\documentclass[conference,a4paper]{IEEEtran}
\IEEEoverridecommandlockouts
\usepackage{cite}
\usepackage{amsmath,amssymb,amsfonts}
\usepackage{algorithmic}
\usepackage{graphicx}
\usepackage{textcomp}
\usepackage{xcolor}

\usepackage[english]{babel}
\usepackage[utf8]{inputenc}

\usepackage{cite}
\usepackage{amsmath,amssymb,amsfonts}
\usepackage{algorithmic}
\usepackage{graphicx}
\usepackage{textcomp}
\usepackage{xcolor}
\usepackage{wrapfig}
\ifCLASSOPTIONcompsoc
    \usepackage[caption=false, font=normalsize, labelfont=sf, textfont=sf]{subfig}
\else
\usepackage[caption=false, font=footnotesize]{subfig}
\fi
\usepackage[english]{babel}
\usepackage{blindtext}
\usepackage{scalerel}
\usepackage{lipsum}
\usepackage{times}
\usepackage{url}
\usepackage{amssymb}
\usepackage{amsfonts}
\usepackage{amsmath}
\usepackage{graphicx}
\usepackage{eurosym}
\usepackage[framed,numbered,autolinebreaks,useliterate]{mcode}
\usepackage{pifont}
\usepackage{enumitem}
\usepackage{amsmath}
\usepackage{amsthm}

\usepackage{mathtools}

\usepackage{graphics}
\usepackage{graphicx}
\usepackage{epsfig,amssymb}
\usepackage{times}
\usepackage{mathrsfs}
\usepackage{array}
\usepackage{amsmath, latexsym, amsfonts, amssymb}
\usepackage[mathscr]{eucal}
\usepackage{array, tabularx}
\usepackage[autostyle]{csquotes}
\usepackage{authblk}
\usepackage{tikz}
\newsavebox{\tempbox}

\usepackage{siunitx}

\usepackage{multirow}
\usepackage{longtable}

\newcommand{\paren}[1]{\left(#1\right)}
\newcommand{\sqparen}[1]{\left[#1\right]}
\newcommand{\brparen}[1]{\left\{#1\right\}}

\def\BibTeX{{\rm B\kern-.05em{\sc i\kern-.025em b}\kern-.08em
    T\kern-.1667em\lower.7ex\hbox{E}\kern-.125emX}}
\begin{document}

\title{The Effect of Multiple Access Categories on the MAC Layer Performance of IEEE 802.11p
\vspace{-10pt}}

\author[$ \ast $]{Geeth P. Wijesiri}
\author[$ \dag $]{Jussi Haapola}
\author[$ \ddag $]{Tharaka Samarasinghe}
\vspace{-0.25cm}
\affil[$ \ast,\ddag $]{Department of Electronic and Telecommunication Engineering, University of Moratuwa, Sri Lanka}
\affil[$ \ast $]{Department of Electrical and Information Engineering, University of Ruhuna, Sri Lanka}
\affil[$\dag $]{Center for Wireless Communications, University of Oulu, Finland}
\affil[$ \ddag $]{Department of Electrical and Electronic Engineering, University of Melbourne, Australia}
    
\affil[$  $ ]{Email: \textit {geeth@eie.ruh.ac.lk, jussi.haapola@oulu.fi, tharakas@uom.lk }\vspace{-18pt} }


\bibliographystyle{ieeetr}
\maketitle

\begin{abstract}
The enhanced distributed channel access (EDCA) mechanism enables IEEE 802.11p to accommodate differential quality of service (QoS) levels in vehicle-to-vehicle (V2V) communications, through four access categories (ACs). This paper presents multi-dimensional discrete-time Markov chain (DTMC) based model to study the effect of parallel operation of the ACs on the medium access control (MAC) layer performance of ITS-G5 IEEE 802.11p. The overall model consists of four queue models with their respective traffic generators, which are appropriately linked with the DTMCs modeling the operation of each AC. Closed-form solutions for the steady-state probabilities of the models are obtained, which are then utilized to derive expressions for key performance indicators at the MAC layer. An application for a highway scenario is presented to draw insights on the performance. The results show how the performance measures vary among ACs according to their priority levels, and emphasize the importance of analytical modeling of the parallel operation of all four ACs. 
\end{abstract}

\begin{IEEEkeywords}
Access categories, discrete time Markov chain, ETSI ITS-G5, IEEE 802.11p, medium access control, vehicle-to-vehicle communication.
\end{IEEEkeywords}
\vspace{-0.4cm}
\section{Introduction}
Vehicular networks primarily depend on vehicle-to-vehicle (V2V) communications for an active safety environment. 
Thus, V2V communications have gained considerable research interest, with 
IEEE 802.11p / dedicated short-range communication (DSRC) as a key enabling technology. 
The enhanced distributed channel access (EDCA) mechanism \cite{edca} was introduced for IEEE 802.11p to allow the vehicle to accommodate differential QoS levels through four access categories (ACs) similar to IEEE 802.11e. The four ACs, namely voice $AC_{vo}$, video $AC_{vi}$, best effort $AC_{be}$ and background $AC_{bk}$, have different parameters for channel contention, such that a vehicle can satisfy the QoS constraints of a multitude of data traffic classes.

European telecommunications standard institute (ETSI) intelligent transportation system (ITS) G5 \cite{b17} defines four traffic classes based on the queues in the access protocol, namely, decentralized environmental notification messages (DENM), high priority DENM (HPD), cooperative environmental notification messages (CAM) and multi-hop DENM (MHD). Each of these queues has an associated AC in IEEE 802.11p, with a defined priority level \cite{b17}. 
This paper studies the impact of parallel multiple ACs on the MAC layer performance of ITS-G5 IEEE 802.11p, by developing a discrete time Markov chain (DTMC) based model. 

The initial work on the MAC layer performance modeling of the IEEE 802.11p EDCA mechanism considered only a selective subset of the four ACs \cite{rp1,rp3,rp4}. 
Subsequently, the authors of \cite{rp2} and \cite{4AC_2} considered the parallel operation of all four ACs in their DTMC based modeling. 
Our work improves these models along multiple facets.
The main novelty in our model is the increased resolution, allowing us to study the whole system performance for each $aSlotTime$, which is the smallest time unit in IEEE 802.11p. This makes our model more complex, but more accurate, and closely resembles the MAC layer performance according to the standard. The increased resolution enables us more precise modeling of the waiting times of different ACs before resuming channel contention. 
As a result, the model correctly captures the effect of prioritization among ACs, and an AC bearing a higher priority can initiate transmission during the longer waiting period of an AC with lower priority. 
In the transmission stage, the transmission delays can be calculated more accurately depending on the size of the payload.
Also, our model improves few slight inconsistencies of \cite{rp2} and \cite{4AC_2} with the standard, in terms of broadcast traffic, {\em e.g.}, \cite{rp2} and \cite{4AC_2} increment the contention window at every attempt the channel is found busy. Note that V2V communications rely on broadcast due to the strict latency constraints. Finally, we consider separate DTMC models for
HPD, DENM, CAM and MHD packet generation, which allows us to alter the traffic arrival patterns among the ACs, and create a more realistic V2V communication scenario, compared to the simple models with predefined traffic rates in \cite{rp2} and \cite{4AC_2}. 


Our contributions can be summarized as follows. We provide detailed DTMC based modeling of the MAC layer protocols of the ETSI ITS-G5 IEEE 802.11p with the parallel operation of multiple ACs. 
The complete model consists of four device level queue models with their respective traffic generators, which are then appropriately connected to the DTMC based state machines that model the operation of each AC. Dependencies are introduced among the models to capture prioritization of different traffic types.  
We obtain closed form expressions for the steady-state probabilities of the DTMCs, which we then use to numerically evaluate MAC layer based performance measures   
such as the average delay, the collision probability, the throughput, and the channel utilization. 
The results show how the performance measures differ among ACs depending on their priority levels and highlight the importance of analytical modeling of the parallel operation of all four ACs. 

The remainder of the paper is organized as follows. The analytical models are presented in Section \ref{sec:secII}. The steady-state solutions and the performance analysis are presented in Sections \ref{sec:secIII} and \ref{sec:secIV}, respectively. The numerical results and discussion follow in Section \ref{sec:secV}, and Section \ref{sec:secVI} concludes the paper.

\vspace{-7pt}
\section{Analytical Models}\label{sec:secII}

The DTMC based modeling of the MAC layer of ITS G5 IEEE 802.11p consists of eleven DTMCs (three for packet generation, four to model the packet queues and four to model the operation of the ACs) that run in parallel. 
All the DTMCs are ergodic, \textit{i.e.,}~they are aperiodic and positive recurrent, hence, a steady-state distribution exists. 
\subsection{The Packet Generators and the Device Level Packet Queues}
The queue models represent the arrival queues of each packet type of interest - CAM (utilized in $AC_{be}$), DENM (utilized in $AC_{vi}$), HPD (utilized in $AC_{vo}$) and MHD (utilized in $AC_{bk}$) \cite{b14}. We skip extensive details of these models due to space limitations. We use the DTMC illustrated in Fig. 1 of \cite{c_v2xvsg5_jp} to model the device level packet queues related to each AC. We use index $i \in \mathcal{AC}= \{vo,vi,be,bk\}$ to differentiate between the states and parameters of the individual queue models related to ACs voice, video, best-effort and background, respectively. 
Each state of the referred queue model is explained by a traffic generator model, depending on the respective AC. We use the DTMC illustrated in Fig. 2 of \cite{c_v2xvsg5_jp} to model the CAM, DENM and HPD packet generators.
CAM packets have a periodic packet arrival pattern, and the inter-arrival time is set according to the standard.
Since HPD and DENM are event triggered, we utilize Poisson arrival processes to determine the arrival process of these packets. We use the same average packet arrival rate of $\lambda$ for both HPD and DENM. Each HPD and DENM packet is repeated $k$ times for reliability. However, to incorporate the higher priority of HPD, we set the repetition interval of two HPD packets to be half of that of two DENM packets.   
For MHD generation, we avoid a DTMC generator model due to its complexity. Thus, we model the packet generation by utilizing a simple Poisson arrival process.
MHD packets are not repeated for added reliability.

\subsection{State Machines for the Four ACs of IEEE 802.11p}
\begin{figure}[t]
    \centering
\includegraphics[scale=.425]{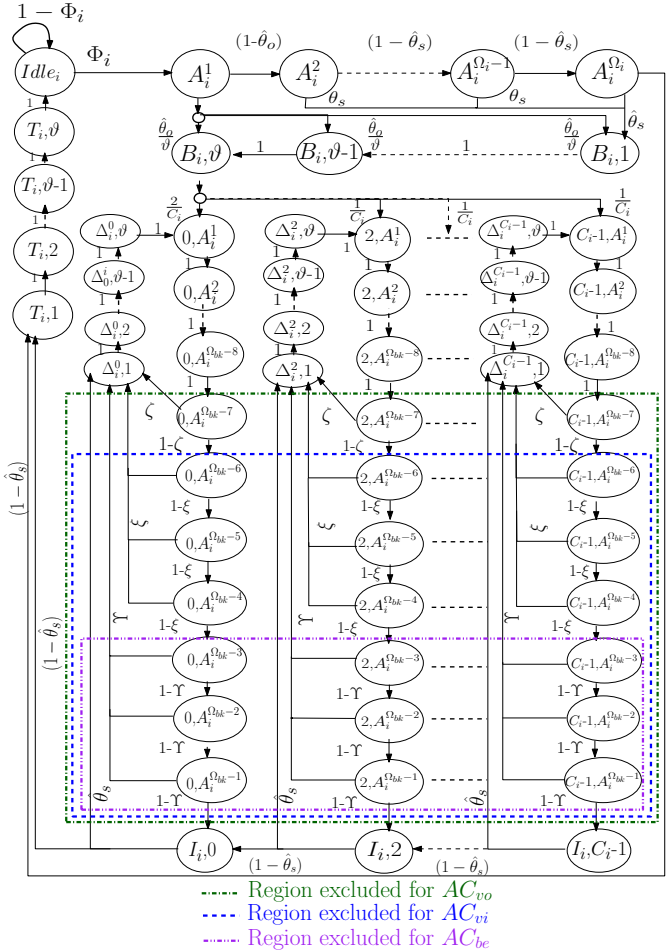}
\vspace{-11pt}
      \caption{DTMC model: State machine for IEEE 802.11p ACs.}
    \label{fig:composite_state_machine}
    \vspace{-0.4cm}
\end{figure}

In this part, we present the DTMCs that are used to model the behavior of each AC of IEEE 802.11p. 
All state machines are developed in a manner that facilitates studying of the MAC layer performance for each $aSlotTime = 13\ \mu$s interval. 
Let $N$ be the number of vehicles in the area of influence, and let $v$ be the target vehicle. 
We use the DTMC model illustrated in Fig. \ref{fig:composite_state_machine} to describe the MAC layer operation of $AC_i$, $i \in \mathcal{AC}$. 

At the start, $v$ is at state $(Idle_{i})$, and remains in this state if the packet queue is empty, or if the 
queues of the ACs of higher priority are non-empty. If a packet arrives when the higher priority queues are empty, $v$ listens for an AIFS duration. The AIFS duration for $AC_i$ denoted by $AIFS_i$ is calculated according to the standard \cite{b16}.
For $AC_i$, let $C_{i}$ denote the minimum contention window size, and $\Omega_{i}=AIFS_i/aSlotTime$.  
States $(A_{i}^j)$ for $ j \in \brparen{1,\dots,\Omega_{bk}}$ represent this waiting time. If the channel stays idle in this duration, the vehicle transmits. Transmission is represented by states $(T_{i},j)$, where $j \in \brparen{1,...,\vartheta }$ and $\vartheta $ denotes the number of $aSlotTime$ intervals required to transmit a packet of 134 bytes over a 6 Mbps control channel (CCH) \cite{b17}.

However, if the channel is found to be occupied within the waiting period, the vehicle waits $\vartheta \times aSlotTime$, which is the time taken for a transmission to end, and the channel to be free again. The wait is represented by states $(B_{i},j)$, where $j \in \brparen{1,\dots,\vartheta}$. The channel being busy at state $(A_{i}^1)$ depicts a scenario where the packet arrival of the vehicle of interest has occurred while the channel is busy, \textit{i.e.}, another vehicle is transmitting. Thus, the time it has to wait until the channel is idle again will be $K \times aSlotTime$, where $K$ is a uniformly distributed random integer in $[1,\vartheta]$. Thus, the transition from state $(A_{i}^1)$ is different to the transitions from $(A_{i}^j)$ for $ j \in \brparen{2,\dots,\Omega_{bk}}$. We can also notice a difference in the transition probabilities. The two transition probabilities represent slight variations of the probability of the channel being busy. At states $(A_{i}^j)$, where $j \in \{2,\dots,\Omega_{bk} \}$, the channel has remained idle for at least one $aSlotTime$ interval. Thus, if $v$ can sense transmission at one of these states, the transmission from a neighboring vehicle must have just initiated as the channel was idle before. Hence, the transmitting device should be at state $(T_{i},1)$. It is not hard to see that this restriction does not exist at $(A_{i}^1)$, and the transmitting device can be at any state $(T_{i},j)$, where $j \in \brparen{1,...,\vartheta }$. Thus, we can expect $\hat{\theta}_{o} \geq \hat{\theta}_{s}$.



After $v$ reaches $(B_{i}, \vartheta)$, where the channel is supposed to be idle again, it initiates a backoff process. A backoff counter value is selected randomly (uniformly) from $\sqparen{0,C_{i}}$, and a backoff stage is selected depending on the selected backoff counter value. According to the standard, backoff counter value 0 and 1 both lead to backoff stage $0$. Thus, the transition probability to backoff stage $0$ is twice the value of any other transition probability. During the backoff, $v$ waits for another AIFS duration before sensing the channel again. For backoff counter value $ b\in \brparen{0,\dots,\paren{C_{i}-1}}$, states $(b, A_{i}^{j})$, where $j \in \brparen{1,\dots, \paren{\Omega_{i}-1}}$, represent this waiting duration.  In the AIFS duration $v$ senses the channel after each $aSlotTime$ interval. If it finds the channel to be busy, it waits for $\vartheta \times aSlotTime$, which is represented by states $(\Delta_{i}^b,j)$, where $j \in  \brparen{1,\dots,\vartheta}$, and another AIFS duration at the same backoff stage.
The same process happens at state $(I_{i}, b)$ if the channel is sensed to be busy. This loop continues until the channel is found to be idle at state $\paren{I_{i},b}$, where the backoff counter is decremented, to arrive at state $(I_{bk},b- 1)$. If $v$ finds the channel to be free at state $(I_{bk},0)$, it transmits its packet. 

The ACs with a higher priority have a shorter AIFS duration, and $AC_{bk}$, which is the AC with the lowest priority, has the longest duration. Thus, Fig. \ref{fig:composite_state_machine} in fact illustrates the DTMC model for $AC_{bk}$. The DTMC models for the other three ACs, that have shorter AIFS duration values, can be obtained by appropriately excluding states and transitions (as illustrated) from Fig. \ref{fig:composite_state_machine}.
For the backoff stage value $b$, the transition from states ($b,A_{i}^{\Omega_{bk}-j}$), where $j\in \{1,\dots,7\}$, to state ($\Delta_{i}^{b}$,1) can only happen due the arrival of a higher priority packet. As an example, for $AC_{bk}$, this can happen due to the arrival of a HPD, DENM or CAM packet. Let $\theta_{i}$ denote the probability of the channel being busy (channel busy ratio) for $AC_i$. Then, $\zeta=\theta_{vo}$, $\xi=(\theta_{vo}+\theta_{vi})$, and $\Upsilon=\sum_{i\in\mathcal{AC}\setminus\{bk\}}\theta_{i}$.



\section{Steady-State Solutions}\label{sec:secIII}
Steady-state solutions of the DTMCs are presented in this section. We denote the steady-state probability of an arbitrary state ($A$) by $\pi_{A}$. Firstly, we obtain expressions for several transition probabilities in  Fig. \ref{fig:composite_state_machine}. To this end, let $P_{arr}^{i}$ and $P_{qe}^{i}$, where $i \in \mathcal{AC}$, denote the packet arrival probability and the probability of the queue being empty for $AC_i$, respectively. It is not hard to show that $\Phi_{vo}=[1-(1-P_{arr}^{vo})P_{qe}^{vo}]$, $\Phi_{vi}  =[1-(1-P_{arr}^{vi})P_{qe}^{vi}]P_{qe}^{vo}$, $\Phi_{be}=[1-(1-P_{arr}^{be})P_{qe}^{be}]P_{qe}^{vo}P_{qe}^{vi}$, and $\Phi_{bk}=[1-(1-P_{arr}^{bk})P_{qe}^{bk}]P_{qe}^{vo}P_{qe}^{vi}P_{qe}^{be}$. 
These probability values can be computed by utilizing the steady-state probabilities of the queue models and the generator models, as shown in \cite{c_v2xvsg5_jp}, which we skip  due to space limitations. Moreover, we approximate $\hat{\theta}_{s}$ and $\hat{\theta}_{o}$ by 
$\hat{\theta}_{s} \approx 1-\sqparen{\prod_{i\in\mathcal{AC}}(1-\pi_{{T_{i}},1})}^{N-1}$ and $\hat{\theta}_{o} \approx 1-\sqparen{\prod_{i\in\mathcal{AC}}\paren{1-\sum_{j=1}^{\vartheta}\pi_{T_{i},j}}}^{N-1}$, respectively, which can be used subsequently to obtain expressions for $\theta_{i}$ for $i \in \mathcal{AC}$. 
For an example, $\theta_{vo}\hspace{-3pt}=\hat{\theta}_{s}\paren{\pi_{A_{vo}^{\Omega_{vo}}}+\pi_{I_{vo},0}}/\left[\sum_{i\in\mathcal{AC}}\paren{\pi_{A_{i}^{\Omega_{i}}}+\pi_{I_{i},0}}\right].$ 
Now that we have all transition probabilities defined, next we obtain expressions for the steady-state probabilities of all states in Fig. \ref{fig:composite_state_machine}. We start by presenting the solutions for the states that are common to all ACs.

\vspace{-5pt}
\subsection{Steady-state Solutions for the States Common to All ACs}
\vspace{-5pt}
  Let $i \in \mathcal{AC}$. We have $\pi_{A_{i}^{j}}=\pi_{idle_{i}}\Phi_{i}$ for $j=1$ and $\pi_{A_{i}^{j}}= \pi_{idle_{i}}\Phi_{i}\paren{\hspace{-2.5pt}1\hspace{-3pt}-\hspace{-3pt}\hat\theta_{o}\hspace{-2.5pt}} \paren{1\hspace{-3pt}-\hspace{-3pt}\hat{\theta}_{s}\hspace{-2.5pt}}^{\paren{j-2}}$ for $j\hspace{-3pt} \in\hspace{-3pt} \sqparen{2, \hspace{-2.5pt} \paren{\hspace{-1.5pt}\Omega_{i}\hspace{-3pt}-\hspace{-3pt}1\hspace{-2.5pt}}}.$
$\pi_{I_{i},j}=\pi_{B_{vo},\vartheta}\paren{C_{vo}-j}/\left[C_{vo}\paren{1-\hat{\theta}_{s}}\right]$ for $ j \in \sqparen{0, \paren{C_{vo}-1}}$. 
$\pi_{B_{i},j}$ $ =\pi_{Idle_{i}}\Phi_{i}\sqparen{j\frac{\hat\theta_{o}}{\vartheta}\hspace{-2pt}+\hspace{-2pt}\paren{1\hspace{-2pt}-\hspace{-2pt}\hat\theta_{o}}\sqparen{1\hspace{-2pt}-\hspace{-2pt}\paren{1\hspace{-2pt}-\hspace{-2pt}\hat{\theta}_{s}}^{\Omega_{i\hspace{-2pt}}-\hspace{-2pt}1}}}\hspace{-2pt}$ and $\pi_{T_{i},j} $ $=\pi_{Idle_{i}}\Phi_{i}$ for $ j \in \sqparen{1, \vartheta}$. 

\vspace{-5pt}
\subsection{Steady-state Solutions for States Specific to $AC_{vo}$}
\vspace{-4pt}
$\pi_{\Delta_{vo}^{b},j}$\hspace{-4pt} =\hspace{-4pt} $\pi_{B_{vo},\vartheta}\paren{C_{vo}-k}\hat{\theta}_{s}/\left[C_{vo}\paren{1-\hat{\theta}_{s}}\right]$ for $b\in \sqparen{0, \paren{C_{vo}-1}}$, $j \in \sqparen{1, \vartheta}$. For $j \in \sqparen{1,  \paren{\Omega_{bk}-8}}$, $\pi_{b,A_{vo}^{j}}\hspace{-4pt}=\hspace{-3pt}\ \pi_{B_{vo},\vartheta}\sqparen{1+\paren{C_{vo}-b- 1}\hat{\theta}_{s}}/\left[C_{vo}\paren{1-\hat{\theta}_{s}}\right] \text{ for } b\in\sqparen{2, \paren{C_{vo}\hspace{-3pt}-\hspace{-3pt}1}} \hspace{-1pt} $ and $\pi_{b,A_{vo}^{j}}=\pi_{B_{vo},\vartheta}\paren{2-2\theta+C_{vo}\hat{\theta}_{s}}/\left[C_{vo}(1 \right.\\ \left.-\hat{\theta}_{s})\right ]\hspace{-2pt}\hspace{-1pt} \text{ for }  b=0 $.
\noindent
\vspace{-8pt}
\subsection{Steady-state Solutions for States Specific to $AC_{vi}$}
\vspace{-5pt}
For $j \in \sqparen{1, \vartheta}, \pi_{\Delta_{vi}^{b},j}= \pi_{B_{vi},\vartheta}[\theta_{vo}+\paren{C_{vi}-b- \theta_{vo}}\hat{\theta}_{s}]\\/\left[C_{vi}(1  -\theta_{vo})(1-\hat{\theta}_{s})\right] \text{ for } b \in [2, \paren{C_{vi}\hspace{-3pt}-\hspace{-3pt}1}] \text{ and } 
\pi_{\Delta_{vi}^{b},j}\hspace{-3pt}=\hspace{-3pt} \pi_{B_{vi},\vartheta}\left[2\theta_{vo}+(C_{vi} -2\theta_{vo})\hat{\theta}_{s}\right]/\left[C_{vi}\paren{1-\theta_{vo}}\paren{1-\hat{\theta}_{s}}\right] \\\text{ for } b =0.$
For $j \in \sqparen{1, \paren{\Omega_{bk}-7}}, \pi_{b,A_{vi}^{j}}\hspace{-3pt}$=$\ \pi_{B_{vi},\vartheta}\left[1+(C_{vi}\right. \\ \left.-b- 1)\hat{\theta}_{s}\right]/\left[C_{vi}(1-\hat{\theta}_{s})\left(1-\theta_{vo}\right)\right] \text{ for }  b  \in  \sqparen{2, \paren{C_{vi}\hspace{-3pt}-\hspace{-3pt}1}}$ and 
$\pi_{b,A_{vi}^{j}}$=$\pi_{B_{vi},\vartheta}\left(2-2\theta+C_{vi}\hat{\theta}_{s}\right) \left[C_{vi}\paren{1-\hat{\theta}_{s}}(1-\theta_{vo})\right]\\ \text{ for } b=0.$
\noindent
\vspace{-5pt}
\subsection{Steady-state Solutions for States Specific to $AC_{be}$}
\vspace{-5pt}
For $j \in \sqparen{1, \vartheta},\ G_1=\theta_{vo}+(1-\theta_{vo})\big[1-(1-\theta_{vo}-\theta_{vi})^3(1-\hat{\theta}_{s})\big], \pi_{\Delta_{be}^{b},j}=\left[G_1+(C_{be}-b- 1)\hat{\theta}_{s}\right]/\left[C_{be}\paren{1-\theta_{vo}}(1-\theta_{vo}-\theta_{vi})^{3} (1\right. \\ \left.-\hat{\theta}_{s})\right]\text{ for } b  \in \sqparen{2, \paren{C_{be}-1}}$ and 
$\pi_{\Delta_{be}^{b},j}$=$\ \left[2G_1+(C_{be}-2)\right. \\ \left.\hat{\theta}_{s} \right]  /\left[C_{be}\paren{1-\theta_{vo}}\paren{1-\theta_{vo}-\theta_{vi}}^{3}(1 -\hat{\theta}_{s})\right] \text{ for } b  =0.$
For $j \in \sqparen{1,(\Omega_{bk}-1)}, G_2(b)\ $=$\ \pi_{B_{be},\vartheta}\sqparen{1+\paren{C_{be}-b- 1}\hat{\theta}_{s}} \\ /\left[C_{be}\paren{1-\hat{\theta}_{s}}\paren{1-\theta_{vo}} (1-\theta_{vo}-\theta_{vi})^{3}\right],$  
\begin{equation}
\hspace{-5pt}f_{1}(j)\hspace{-4pt}=\hspace{-3pt}
\begin{dcases}
1/(1-\theta_{vo}) \text{ for } j \in [1,(\Omega_{bk}\hspace{-2pt}-\hspace{-2pt}4)]\\
(1\hspace{-2pt}-\hspace{-2pt}\theta_{vo}\hspace{-2pt}-\hspace{-2pt}\theta_{vi})^{b}\text{ for } j \hspace{-2pt}=\hspace{-2pt}(\Omega_{bk}\hspace{-2pt}-\hspace{-2pt}3\hspace{-2pt}+\hspace{-2pt}b)\text{ and } b\in \brparen{0,1,2}
\end{dcases}.\nonumber
\end{equation}
$\pi_{b, A_{be}^{j}}$=$\ G_{2}(b)f_1(j)(1-\theta_{vo}) \text{ for } b  \in \sqparen{2, C_{be}-1}$, and
$\pi_{b, A_{be}^{j}}=G_{2}(b)[2\hspace{-2.5pt}+\hspace{-2.5pt}(C_{be}\hspace{-2.5pt}-\hspace{-2.5pt}2)\hat{\theta}_{s}]/[1\hspace{-2.5pt}+\hspace{-2.5pt}(C_{be}\hspace{-2.5pt}-\hspace{-2.5pt}b\hspace{-2.5pt}-\hspace{-2.5pt}1) \hat{\theta}_{s}] \text{ for }b=0.$ 

\noindent

\subsection{Steady-state Solutions for States Specific to $AC_{bk}$}
For $j \in \sqparen{1, \vartheta},  G_3\ $=$\ \theta_{vo}+(1-\theta_{vo})\big[1-(1-\theta_{vo}-\theta_{vi})^3[(1-\sum_{l\in\mathcal{AC}\setminus\{bk\}}\theta_{l})^{3}+(1-\hat{\theta}_{s})]\big], \pi_{\Delta_{bk}^{b},j} \hspace{-2.5pt}=\hspace{-2.5pt}\left[G_3+(C_{bk}-b-1)\hat{\theta}_{s}\right]/\left[{C_{bk}(1-\theta_{vo}}(1-\theta_{vo}-\theta_{vi})^{3}(1
\right. \\ \left.-\sum_{l\in\mathcal{AC} \setminus \{bk\}}\theta_{l})^{3}(1-\hat{\theta}_{s})\right]\text{for } b  \in\ [2, (C_{bk}\hspace{-2.5pt}-\hspace{-2.5pt}1)]\text{ and } \pi_{\Delta_{bk}^{b},j}\\=\left[2G_3+(C_{bk}-2)\hat{\theta}_{s}\right]/\left[C_{bk}(1-\theta_{vo})(1-\theta_{vo}-\theta_{vi})^{3}(1\right. \\ \left.-\theta_{vo}-\theta_{vi}-\theta_{be})^{3}\paren{1\hspace{-2.5pt}-\hspace{-2.5pt}\hat{\theta}_{s}}\right]\text{for } b\hspace{-2.5pt} =\hspace{-2.5pt}0.$
For $j\in[1,(\Omega_{bk}-1)],\\ G_4(b)= \pi_{B_{bk},\vartheta}[1+\paren{C_{bk}-b\hspace{-3pt}-1}\hat{\theta}_{s}]/\left[C_{bk}(1-\hat{\theta}_{s})(1-\theta_{vo})\right. \\ \left.(1-\theta_{vo}-\theta_{vi})^{3}(1-\sum_{l\in\mathcal{AC}\setminus\{bk\}}\theta_{l})^{3}\right],$
\begin{equation}
\hspace{-5pt}f_{2,j}\hspace{-4pt}=\hspace{-3pt}
\begin{dcases}
1/(1-\theta_{vo}) \text{ for } j \in [1,(\Omega_{bk}\hspace{-3pt}-\hspace{-3pt}7)]\\
(1\hspace{-3pt}-\hspace{-3pt}\theta_{vo}\hspace{-3pt}-\hspace{-3pt}\theta_{vi})^b\text{ for } j \hspace{-3pt}=\hspace{-3pt}(\Omega_{bk}\hspace{-3pt}-\hspace{-3pt}6\hspace{-3pt}+\hspace{-3pt}b) \text{ and } b  \in \brparen{0,1,2,3}\\
(1\hspace{-3pt}-\hspace{-3pt}\theta_{vo}\hspace{-3pt}-\hspace{-3pt}\theta_{vi})^3(1\hspace{-3pt}-\hspace{-3pt}\theta_{vo}\hspace{-3pt}-\hspace{-3pt}\theta_{vi}\hspace{-3pt}-\hspace{-3pt}\theta_{be})^{b+1}\\\hspace{80pt} \text{for } j\hspace{-3pt} =\hspace{-3pt}(\Omega_{bk}\hspace{-3pt}-\hspace{-3pt}2+b)\text{ and } b\in\brparen{0,1}\\
\end{dcases}.
\nonumber
\end{equation}
$\pi_{b, A_{bk}^{j}}=G_4(b) \times f_{2}(j) \text{ for } b  \in \sqparen{2, C_{bk}-1}\text{ and }\ \pi_{b, A_{bk}^{j}}\\=f_{2}(j)(1-\theta_{vo})G_4(b)[2+(C_{bk}-2)\hat{\theta}_{s}]/\left[1+(C_{bk}-b\right.\\\left.-1)\hat{\theta}_{s}\right] \text{for }  b=0.$

\noindent


By using the sum of steady-state probabilities, and appropriately substituting the above derived steady-state probabilities, we can obtain $\pi_{idle_{i}}$ for $i \in \mathcal{AC}$, which can then be used to calculate all steady-state probabilities of interest.

\section{Performance Analysis}\label{sec:secIV}
In this section, we derive expressions that can be utilized to study the MAC-layer performance of IEEE 802.11p ACs. In deriving the expressions, we assume the independence among the DTMCs representing the four ACs for a particular vehicle when required (although they are correlated), for mathematical tractability.

\subsection{Probability of Collision }
A collision occurs when two or more vehicles initiate their transmission simultaneously, that is they arrive at state ($T_i,1$) simultaneously. 
Thus, the instantaneous collision probability per $aSlotTime$ can be obtained by eliminating the probabilities of no one initiating transmission or exactly one initiating transmission, and is given by
$P_{col}^{tot}=1-\sqparen{\prod_{i\in\mathcal{AC}}(1-\pi_{T_{i},1})}^{N}-N\sum_{i\in\mathcal{AC}}(\pi_{T_{i},1}\theta_i)\sqparen{\prod_{j\in\mathcal{AC}}(1-\pi_{T_{j},1)}}^{N-1}$.

\subsection{Average Delay}
For $AC_i$, $i\in\mathcal{AC}$, the cycle time of state ($T_{i},1$) can be written as $aSlotTime / \pi_{T_{i},1}$, which gives the average time taken to initiate two consecutive transmissions. 
Note that, the calculated cycle time also includes the time spent by $v$ at state ($idle_i$) without any packets to transmit. As this cannot be considered a part of the delay, we scale the cycle time such that we can eliminate the effect of being idle on the calculated delay.  The resulting modified cycle time for $AC_i$ is given by $\psi_i=(1-P_I^i) \times aSlotTime / \pi_{T_{i},1}$, where $P_I^i$ denotes the probability of $v$ being idle with an empty packet queue for $AC_i$. 
To this end, $P_{I}^{vo}=\pi_{idle_{vo}}$ as the AC with the highest priority can only be in the idle state due to having an empty queue. For other ACs, $v$ can be in the idle state due to its own queue being empty or the queues of the higher priority ACs being nonempty. Therefore, $P_{I}^i$, for $i\in\mathcal{AC}\setminus\{vo\}$ is obtained by multiplying $\pi_{idle_i}$ by $P_{qe}^{i}/P_{idle_i}$, where $P_{idle_i}$ denotes the probability of being in state $(idle_i)$.  
Thus, $P_{I}^{vi}=\pi_{idle_{vi}}P_{qe}^{vi}/[1-(1-P_{qe}^{vi})P_{qe}^{vo}]$, $P_{I}^{be}=\pi_{idle_{be}}P_{qe}^{be}/[1-(1-P_{qe}^{be})P_{qe}^{vo}P_{qe}^{vi}]$ and  $P_{I}^{bk}=\pi_{idle_{bk}}P_{qe}^{bk}/[1-(1-P_{qe}^{bk})P_{qe}^{vo}P_{qe}^{vi}P_{qe}^{be}]$.  

The maximum delay for a packet at the front of a queue for $AC_i$ is $\psi_{i} + (\vartheta-1)aSlotTime$. We have added  $(\vartheta-1)aSlotTime$ to the modified cycle time as the packet arrival at the front of the queue may happen simultaneously with the initiation of transmission of the previous packet, and we are considering the maximum delay. 
Since we consider a queue of length $m$, we obtain an average delay by averaging over the steady-state probability of each state in the queue model in Fig. 1 of \cite{c_v2xvsg5_jp}. Thus, the average delay is given by $D_{i}=[\psi_{i} + (\vartheta-1)aSlotTime] \sum_{j=0}^{m}(j+1)\pi_{j}^{i}$, where $\pi_{j}^{i}$ denotes the steady-state probability of state $j$ of the queue model for $AC_{i}, i \in \mathcal{AC}$.

\subsection{Throughput and channel utilization}
The throughput of each $AC_i$, $i\in\mathcal{AC}$, is calculated through the product between the bandwidth of the CCH and the probability of exactly one vehicle operating in that particular AC transmitting. 
Thus, the throughput for $AC_i$ is given by $S_{i}=\text{bandwidth of CCH}\times N\sum_{j=1}^{\vartheta} \pi_{T_{i},j}\sqparen{\prod_{i\in\mathcal{AC}}(1-\sum_{j=1}^{\vartheta} \pi_{T_{i},j})}^{(N-1)}$. 
Along the same lines, the total throughput denoted by $S_{tot}$, which is calculated by the product of the bandwidth of the CCH and the probability of exactly one vehicle operating in any AC transmitting.
Thus, $S_{tot}=\text{bandwidth of CCH}\times N\sum_{i\in\mathcal{AC}}\paren{\sum_{j=1}^{\vartheta}\pi_{T_{i},j}\theta_i}\sqparen{\prod_{i\in\mathcal{AC}}(1-\sum_{j=1}^{\vartheta} \pi_{T_{i},j})}^{(N-1)}.$
The channel utilization captures the probability of at least one vehicle transmitting. Thus,
$CU =1 - [\prod_{i\in\mathcal{AC}}(1- \sum_{j=1}^{\vartheta} \pi_{T_{i},j})]^{N}.$

\section{Numerical Results and Discussion}\label{sec:secV}

In this section, we present an application of the models for a highway scenario to provide insights and comparisons on key performance indicators, through numerical evaluations. 
We also compare the results with similar ones generated according to the DTMC model in \cite{c_v2xvsg5_jp} (where only one AC is considered) to further highlight the importance of studying a model where all ACs operate in parallel.

We consider a highway with four parallel lanes in each direction. We assume that the vehicles move at a constant speed while keeping the inter vehicle gap according to the two-second rule. We consider that HPD, DENM, CAM, and MHD are utilized for V2V communication \cite{b14, b15}, and their reference packet formats are specified according to \cite{b13}. Vehicle $v$ broadcasts all packets to the $N-1$ neighboring vehicles in the geographical region of influence. The maximum size of the queue models is considered to be 10 packets and packet repetition rate $k$ is taken as 5. The CAM packets inter-arrival time is set at $100$ ms according to \cite{b13}. The average packet inter-arrival rate of the event triggered HPD and DENM, is set at 1 packet per second, whereas the average packet inter-arrival rate of MHD packets is set at 10 packets per second. 

With these parameters, the steady-state solutions of the packet generators, the queue models, and the state machines modeling the ACs, are calculated iteratively in parallel. In the iterative process, we first solve the generators of each AC after initializing values as done in \cite{c_v2xvsg5_jp}. The steady-state probabilities of the generator models are appropriately set as the transition probability values of the queue models. This allows us to solve the queue models. 
We calculate $P_{qe}^{i}$ and $P_{arr}^{i}$ for each $i \in \mathcal{AC}$ from the steady-state probabilities of the respective queue and generator models. These values are then used to solve the DTMCs that model the ACs, such that for $AC_i$, we can calculate values for $\theta_i$ and $P_{s}^i=\sum_{j=1}^{\vartheta}\pi_{T_{i},j}$, which denotes the probability of successful transmission of a packet for $AC_i$. The value of $P_{s}^i$ is used to update the generator models again, and we continue the iterative process until the values for $P_{qe}^{i}$, $\theta_i$, and $P_{s}^{i}$ converge for all $i \in \mathcal{AC}$. 

\subsubsection{Probability of Collision}\label{Sec:B_1}
\begin{figure}[!t]
  \includegraphics[scale=.59]{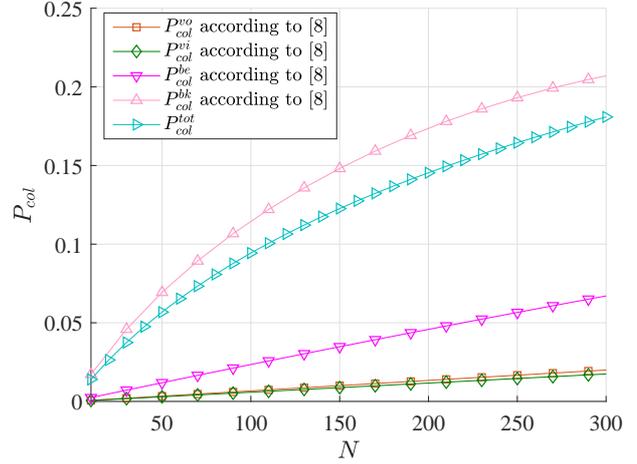}
  \vspace{-10pt}
  \caption{The behavior of the collision probability with the number of vehicles.}
  \label{fig:P_col_vs_N}
  \vspace{-.25cm}
\end{figure}

The behavior of the collision probability with $N$ is illustrated in Fig. \ref{fig:P_col_vs_N}. The collision probability monotonically increases with $N$ and reaches approximately 18\% when $N=300$. 
We can observe that the model in this paper shows that the actual collision probability in a network is much higher compared to what was exhibited through DTMC based models that considered the operation of only one AC. Specifically, this is true for  $AC_{vo}$, $AC_{vi}$ and $AC_{be}$. However, we can observe it to be lower compared to a scenario where only $AC_{bk}$ is considered. The system uses $AC_{bk}$ to transmit MHD packets, which has a higher packet arrival rate compared to the other three ACs. Thus, the collision probability is higher. 
However, in a more realistic setting, all vehicles do not transmit MHD packets simultaneously, leading to a reduction in the total number of packets transmitted in the network, thus a lower collision probability, {\em i.e.}, $P_{col}^{tot}$. We can expect to observe higher collision probability results if same packet queues were fed to the models in \cite{rp2} and \cite{4AC_2}. \cite{rp2} decrements the back-off counter regardless of the channel being busy, thus getting more frequent transmission opportunities. Both \cite{rp2} and \cite{4AC_2} skip the $AIFS$ waiting interval before resuming contention. This restricts them from differentiating among the priority levels of the ACs with regards to channel contention, leading to more frequent transmissions as well. The impact is expected to be higher for high priority ACs.


\subsubsection{Average Delay}
\begin{figure}[t]
  \includegraphics[scale=.59]{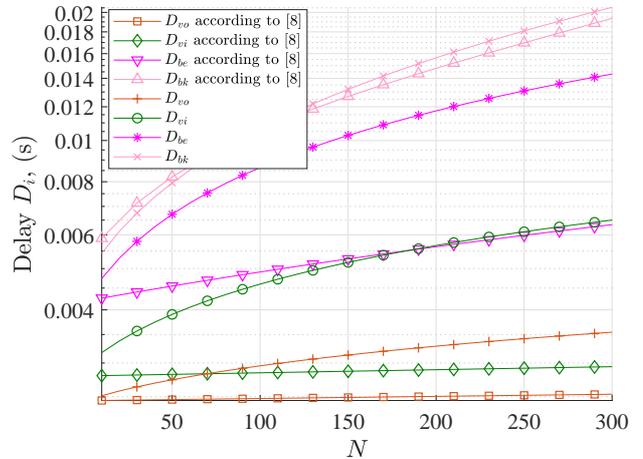}
  \vspace{-10pt}
  \caption{The behavior of the delay with the number of vehicles.}
  \label{fig:delay_vs_N}
  \vspace{-.3cm}
\end{figure}

Fig. \ref{fig:delay_vs_N} illustrates the variation of the delay with $N$. We can observe the lower priority ACs encountering a higher delay compared to the higher priority ACs, as expected. 
The higher priority ACs use shorter CW sizes and AIFS duration values leading to smaller delay values. Also, they get prioritized over the others upon transmission. Similar to the collision probability, we can observe that the delay values are much greater when a model considers all ACs simultaneously, which is indeed the case in a realistic system. 
We can also observe that the parallel operation has higher impact on the delay values of the ACs with lower priorities. This is due to them getting the least preference upon transmission. 
Due to this adverse behavior, it is interesting to study the performance of lower priority classes when the traffic load of higher priority classes increases further. To this end, we increase $\lambda$ and $k$ of the HPD and DENM generators to 10 each and evaluate the service time of CAM packets. The service time can be found to be 7.84 ms at $N=50$ and 16.68 ms at $N=300$, which is well below the lowest inter-arrival time between two CAM packets of 100 ms, according to the standard.
The average delay results obtained according \cite{rp2} and \cite{4AC_2} with the same input queues are expected to be lower as the $AIFS$ waiting intervals before resuming contention are neglected, and \cite{rp2} decrements its backoff counter regardless of the channel being busy, thus getting a transmit opportunity sooner. 


\subsubsection{Throughput and Channel Utilization}
\begin{figure}[t]
  \includegraphics[scale=.59]{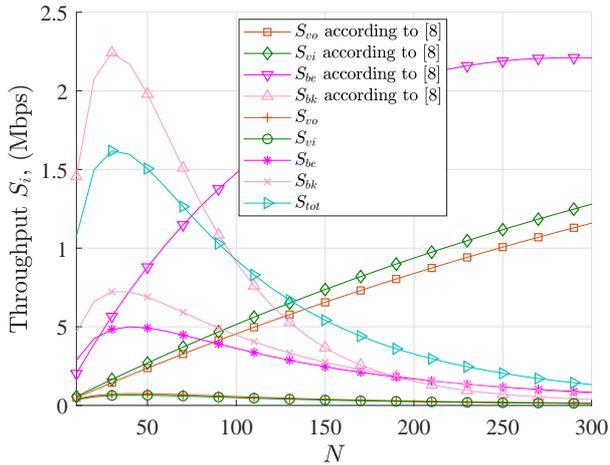}
  \vspace{-10pt}
  \caption{The behavior of throughput with the number of vehicles.}
  \label{fig:TH_vs_N}
  \vspace{-0.25cm}
\end{figure}

Fig. \ref{fig:TH_vs_N} illustrates the behavior of the throughput with $N$. We can observe the throughput increasing with $N$ first, which is rather intuitive. However, when more vehicles contend for transmission simultaneously, the delays and the collisions increase, and hence, the throughput starts to decline. 
The value of $N$ at which the decline starts reduces when the throughput increases, as expected. It is interesting to note that there is a large disparity between the model in this paper compared to \cite{c_v2xvsg5_jp} in terms of throughput. We can see that for all except one, the throughput curves according to \cite{c_v2xvsg5_jp} are increasing even at $N=300$, where as in reality, the value of $N$ that allows the peak throughput is rather small. It is approximately around 30 according to the total throughput curve $S_{tot}$.    
We can study the steady-state probabilities of the queues for further insights on this phenomenon. 
When $N= 30$, the MHD queue saturates (queue is full), and the queue, which has the lowest priority, starts to drop the MHD packets. We can observe the queues getting saturated in the order of their priority when $N$ increases further. These results are not included due to space limitations.  

\begin{figure}[t]
  \includegraphics[scale=.59]{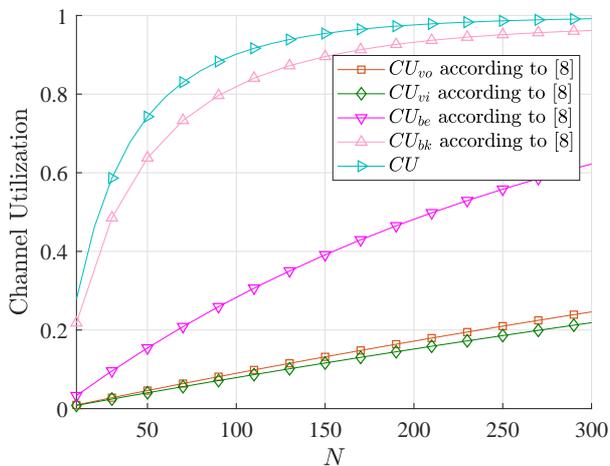}
  \vspace{-10pt}
  \caption{The behavior of channel utilization with the number of vehicles.}
  \label{fig:CU_vs_N}
  \vspace{-0.25cm}
\end{figure}

Fig. \ref{fig:CU_vs_N} illustrates the behavior of the channel utilization with $N$, and the behavior can be explained using similar reasoning as above. 
It is interesting to note that the parallel operation makes best use of the channel with a value of 99.22\% at $N=300$.


\vspace{-5pt}
\section{Conclusions}\label{sec:secVI}
This paper has presented multi-dimensional DTMCs that model the parallel operation of multiple ACs in IEEE 802.11p, considering HPD, DENM, CAM and MHD packets. The packet generation and queues have been appropriately modeled utilizing DTMCs, and have been coupled with the DTMCs that model the operation of ACs. The steady-state probabilities have been obtained in closed-form, and they have been utilized to derive expressions for key performance metrics at the MAC layer. A highway scenario has been considered for numerical results. The numerical results have shown how the high priority queues experience a lower delay and collision probability values compared to the lower priority queues. Furthermore, the results have shown how the throughput firstly increases and then decreases with the number of vehicles in the network. Also, a comparison with the performance achieved when only a single AC is in operation has been presented to highlight the importance of the parallel AC model in the paper.
\vspace{-10pt}
\section*{Acknowledgment}
\vspace{-5pt}
This research has been partially financially supported by the ITEA3 project APPSTACLE (15017) and Academy of Finland 6Genesis Flagship (grant 318927).
\ifCLASSOPTIONcaptionsoff
  \newpage
\fi
\vspace{-5pt}
\bibliography{Geeth-bibfile}

\end{document}